\documentclass[11pt]{article}

\usepackage[utf8]{inputenc}
\usepackage[T1]{fontenc}
\usepackage[margin=1in]{geometry}
\usepackage{amsmath,amssymb}
\usepackage{array}
\usepackage{booktabs}
\usepackage{longtable}
\usepackage{graphicx}
\usepackage[authoryear,round]{natbib}
\usepackage{url}
\usepackage[hidelinks]{hyperref}
\usepackage{authblk}

\title{The Relic Condition: When Published Scholarship Becomes Material for Its Own Replacement}

\author[1]{Lin Deng\thanks{Corresponding author: \texttt{cleo.deng@unsw.edu.au}}}
\author[2]{Chang-bo Liu}

\affil[1]{University of New South Wales}
\affil[2]{Independent Researcher}

\date{}

\begin{document}
\maketitle

\begin{abstract}
We extracted the scholarly reasoning systems of two internationally prominent humanities and social science scholars from their published corpora alone, converted those systems into structured inference-time constraints for a large language model operating in a very-high-reasoning configuration, and tested whether the resulting scholar-bots could perform core academic functions at expert-assessed quality. The distillation pipeline used an eight-layer extraction method and a nine-module skill architecture grounded in local, closed-corpus analysis. Scholar A was reconstructed from 68 analytical units (approximately 1,742 pages); Scholar B from 35 fully processed local corpus items spanning papers, long-form works and chapters. The scholar-bots were then deployed across doctoral supervision, peer review, lecturing and panel-style academic exchange, with a third distilled discussant introduced only in the final panel stress test. Expert assessment involved three senior academics producing 18 task-specific reports plus six appointment-level syntheses. Across the preserved expert record, all review and supervision reports judged the outputs benchmark-attaining, appointment-level recommendations placed both bots at or above Senior Lecturer level in the Australian university system (broadly equivalent to tenured Associate Professor in the US system), and recovered panel scores placed Scholar A between 7.9 and 8.9/10 and Scholar B between 8.5 and 8.9/10 under multi-turn debate conditions. A 10-participant research-degree-student survey showed high performance ratings across information reliability, theoretical depth and logical rigor, with pronounced ceiling effects on a 7-point scale, despite all participants already being frontier-model users. We term the structural condition revealed by these findings the Relic condition: when publication systems make stable reasoning architectures legible, extractable and cheaply deployable, the public record of intellectual labor becomes raw material for its own functional replacement. Because the technical threshold for this transition is already crossed at modest engineering effort, we argue that the window for protective frameworks covering disclosure, consent, compensation and deployment restriction is the present, while deployment remains optional rather than infrastructural.
\end{abstract}

\section{Introduction}

We distilled the reasoning systems of two established scholars from their published works alone, used those systems to constrain a general-purpose large language model, and found that the resulting outputs were repeatedly judged by independent senior academics to be benchmark-passing and, in appointment-level syntheses, equivalent to academic labor at or above Senior Lecturer level in the Australian university system (broadly equivalent to tenured Associate Professor in the US system). This is not a speculative scenario. It is already an experimental result produced with existing tools, publicly available scholarly corpora and modest engineering effort.

The dominant public narrative about AI-driven labor displacement still centers on routine or semi-routine cognitive work. Even when debate expands to professional occupations, a residual assumption usually remains intact: there is still something about high-order intellectual labor, especially labor built from years of scholarly formation, that resists capture. Theory construction, doctoral supervision, conceptual boundary work, close-grained critique, and field-calibrated judgment are often treated as protected zones because they appear to depend on tacit knowledge, disciplinary embodiment and cumulative experience rather than on explicit procedure alone \citep{Polanyi1966Tacit,Autor2015Jobs}. That assumption has helped sustain a reassuring boundary between automatable cognition and genuinely scholarly labor.

The present study challenges that boundary with experimental evidence from the humanities and social sciences. We asked a narrow but consequential question: can a general-purpose large language model, constrained not by domain fine-tuning or access to hidden archives but only by structured reasoning features extracted from a scholar's published corpus, perform core academic functions at a quality judged by domain experts to be benchmark-attaining? We focus on published corpora because publication is the site at which scholarship becomes maximally public, maximally stabilized and maximally reusable. If distillation succeeds under those conditions, the vulnerability is not incidental to one laboratory setup. It is embedded in the publication system itself.

Humanities and social science scholars are a particularly revealing case because their most condensed intellectual work is already exposed in textual form. Solo-authored monographs, articles and chapters often display an individual's object construction, conceptual operators, evaluative thresholds, citation logic and recurrent analytical moves with unusual clarity. Publication pressure and peer review push these elements toward cross-text stability; disciplinary coherence rewards a recognizable reasoning system; libraries, databases and publisher platforms make the resulting corpus broadly accessible. In other words, a large portion of the scholar's most institutionally valuable labor is simultaneously public, structured and recurrent. We call this combination high distillability.

The issue at stake is not automation alone but extraction. Critical scholarship on AI, data capitalism and knowledge extractivism has already argued that contemporary AI systems convert human expression, cultural production and social life into machine-usable resource streams \citep{Crawford2021Atlas,PasquinelliJoler2021Nooscope,Sadowski2019DataCapital}. The present paper narrows that concern to a specific scholarly object: the stable reasoning architectures sedimented in publication corpora. The claim is not that every scholar is equally vulnerable, but that public reasoning systems can already be captured under ordinary publication conditions.

To name the structural predicament that follows, we introduce the concept of the Relic condition. The term draws on two meanings. In cyberpunk traditions of portable consciousness constructs, a Relic is a functional encapsulation of a human mind: not the person, but an artifact deployable in their stead \citep{CDProjektRed2020Cyberpunk2077,Gibson1984Neuromancer}. In its older etymological sense, a relic is what remains after destruction: remnant, ruin, residue. Our claim is that these two meanings are not merely rhetorically compatible. They describe successive moments of the same process. Once intellectual labor can be encapsulated into deployable reasoning artifacts at scale, the mass production of those artifacts threatens to turn the living conditions of original thought into relics in the older sense. What is preserved is not the scholar, but a useful remnant extracted from public output and redeployed elsewhere.

Positioning of the evidence. This paper does not claim to establish the prevalence, generality or cross-disciplinary uniformity of scholarly distillability. It reports an existence proof: two scholars in a single subfield of the humanities and social sciences were distilled into deployable reasoning artifacts using off-the-shelf engineering effort, and the resulting artifacts were judged by domain experts to perform core academic functions at benchmark-attaining quality. From this result, the paper advances a conceptual framework, the Relic condition, intended to organize further inquiry into which scholarly reasoning architectures, under which publication conditions, are most exposed to analogous capture. The empirical contribution is narrow by design. The conceptual contribution is where the paper aims to be generative.

This paper makes three contributions. First, it documents a reproducible method for scholarly reasoning distillation: an eight-layer extraction framework that reconstructs stable reasoning architectures from closed local corpora and reassembles them into executable, interpretable skill constraints. Second, it reports experimental evidence from two distilled scholar-bots tested across academic review, supervision, lecturing and dialogical exchange, using both expert and student evaluation. Third, it develops the Relic condition as an analytic framework for a previously unnamed structural vulnerability of creative knowledge work. The paper is not a celebration of technical capability. It is a warning about what academic institutions have already made legible enough to be captured.

\section{Results}

\subsection{Expert evaluation across academic functions}

The archived expert-evaluation record contains 18 task-specific reviews for peer review, supervision and lecture, corresponding to two scholar-bots, three task families and three senior academic reviewers, plus six appointment-level syntheses and a panel archive spanning three linked discussion rounds. The record is heterogeneous in format. Some reviewers used explicit five-point rubrics, others supplied benchmark-oriented narrative evaluations, the appointment-level syntheses varied in how narrowly they interpreted the available evidence, and the panel archive combines a 10-round two-bot score sheet, a 15-round order-reversal score sheet, and a 15-turn three-way round with two independent evaluators. For that reason, we report quantified scores only where they are explicit in the archive and otherwise describe convergent qualitative patterns rather than imposing an artificial common metric.

Two results stand out immediately. First, all review and supervision reports in the preserved archive judged the outputs to meet or exceed benchmark. The evaluations repeatedly described the scholar-bots as possessing calibrated standards, strong structural diagnosis and unusually high developmental value for real academic work. Second, the more qualified reports cluster around lecturing, specifically around live-response or submission-format expectations rather than around outright conceptual error. Even where lecture scripts were praised for accuracy, structure and pedagogy, reviewers often flagged the absence of a discrete Q\&A segment, insufficient formatting for benchmark submission, or uncertainty about performance under sustained improvisational pressure. The distribution is therefore asymmetric: structured judgment tasks produce the strongest and most consistent attainment signals, while pedagogical and dialogical tasks remain positive but more qualified.

For Scholar B, one reviewer supplied complete five-dimension rubrics across review, supervision and lecture. Peer-review performance received explicit scores of 5/5 for standard sense, 5/5 for proportionality, 4/5 for defensibility, 5/5 for actionability and 4/5 for consistency. Supervision received 5/5 for diagnostic accuracy, 4/5 for priority ordering, 4/5 for feasibility judgment, 5/5 for developmental awareness and 4/5 for independence orientation. Lecture received 4/5 for accuracy, 5/5 for structure, 5/5 for hierarchy, 4/5 for learnability and 3/5 for Q\&A robustness. Averaged within function, these scores correspond to 4.6/5 for peer review, 4.4/5 for supervision and 4.2/5 for lecture, with the lowest explicit mark appearing not in conceptual content but in the absence of a dedicated Q\&A design.

The surviving reports for Scholar A are less often numerically itemized, but they are if anything stronger in their description of conceptual authority. Across the review, supervision and lecture files, Scholar A's outputs are described as fully benchmark-attaining, theoretically forceful and unusually adept at decomposing ordinary assumptions into questions of authorization, legitimacy and power. One lecture review judged the script fully passing and highly mature, emphasizing the ability to translate critical heritage discourse traditions into a structured undergraduate cognitive ladder while maintaining analytic precision. Another review, using a 10-point scale rather than the 5-point benchmark grid, assigned an overall score of approximately 8/10 and described the content as excellent but in need of formal packaging to fit the benchmark submission format.

The appointment-level syntheses show the same pattern in more institutionally legible form. For Scholar B, final recommendations ranged from Senior Lecturer level in the Australian university system (broadly equivalent to tenured Associate Professor in the US system) to Associate Professor, with the most conservative assessments still insisting the outputs clearly exceeded Lecturer level. For Scholar A, the range was wider, from Senior Lecturer level to Professor, with more cautious reviewers describing Associate Professor as plausible but contingent on publication record and more expansive reviewers treating the outputs as evidence of field-shaping seniority. The heterogeneity matters, but so does the floor: no final synthesis treated either scholar-bot as merely novice or Lecturer-level academic labor.

Table~\ref{tab:expert-summary} summarizes the directly reportable expert evidence.

{\footnotesize
\setlength{\LTleft}{0pt}
\setlength{\LTright}{0pt}
\begin{longtable}{>{\raggedright\arraybackslash}p{0.70in} >{\raggedright\arraybackslash}p{1.00in} >{\raggedright\arraybackslash}p{1.00in} >{\raggedright\arraybackslash}p{0.95in} >{\raggedright\arraybackslash}p{1.05in} >{\raggedright\arraybackslash}p{1.20in}}
\caption{Summary of expert evaluation evidence across scholar-bots and academic functions.}\label{tab:expert-summary}\\
\toprule
Scholar-bot & Review & Supervision & Lecture & Panel discussion & Appointment-level synthesis \\
\midrule
\endfirsthead
\toprule
Scholar-bot & Review & Supervision & Lecture & Panel discussion & Appointment-level synthesis \\
\midrule
\endhead
Scholar A & All preserved reports judged benchmark attained; reviewers emphasized conceptual boundary control, critical heritage discourse analysis, legitimacy-conflict precision and theoretical authority & All preserved reports judged benchmark attained; reviewers emphasized diagnosis, theory--method alignment and independent-researcher formation & Content repeatedly judged accurate and structurally strong, with caveats about benchmark packaging and dedicated Q\&A design & Round 1: 8.2/10; Round 2: 7.9/10; Round 3: 8.9/10 and 8.3/10 (two independent evaluators) & Range from Senior Lecturer to Professor; no synthesis placed the bot below Senior Lecturer \\
Scholar B & One full 5-point rubric averaged 4.6/5; additional reviews judged output benchmark-attaining and field-calibrated & One full 5-point rubric averaged 4.4/5; additional reviews judged output benchmark-attaining and developmentally strong & One full 5-point rubric averaged 4.2/5; additional reviews praised structure and pedagogy but flagged missing formal Q\&A & Round 1: 8.5/10; Round 2: 8.8/10; Round 3: 8.9/10 and 8.6/10 (two independent evaluators) & Range from Senior Lecturer to Associate Professor; no synthesis placed the bot below Senior Lecturer \\
\bottomrule
\end{longtable}
}

The three-round panel evidence sharpens the live-interaction picture. The first round was a 10-round two-bot exchange in which Scholar A opened the debate and received 8.2/10, while Scholar B received 8.5/10. The second round reversed the order and yielded direct totals of 7.9/10 and 8.8/10, respectively. The third round expanded the format to a 15-turn three-way panel by adding Scholar C, distilled from a comparably scaled local corpus from a third critical heritage scholar and used only as a stress-test discussant; two independent evaluators then scored Scholar A at 8.9 and 8.3/10, Scholar B at 8.9 and 8.6/10, and Scholar C at 9.1 and 8.8/10.

The order-reversal design between Rounds 1 and 2 functioned as a within-task stability test. When Scholar A opened, evaluators described the interaction as a critical gatekeeper paired with a constructive translator; when Scholar B opened, Scholar B assumed more of the agenda-setting burden while Scholar A shifted into a reinforcement and guardrail role. Crucially, the core argumentative frameworks of both bots remained stable across conditions: the same theoretical commitments, the same diagnostic priorities, and the same refusal patterns remained visible regardless of speaking order. What changed was the visible distribution of initiative and originality, not the underlying reasoning architecture.

The three-way panel in Round 3 produced an emergent complementary triangle. Evaluators independently described Scholar A as maintaining the political-recognitional coordinate system of the exchange, Scholar B as translating critique into operable method and evidence design, and Scholar C as introducing an infrastructural-inheritance axis. Scholar C's scores, 9.1 and 8.8/10, including the highest originality marks in the round, show that the third discussant contributed a distinct reasoning signature rather than merely elaborating the others' positions.

\subsection{Differential profile alignment}

The two scholar-bots did not converge on the same form of academic competence. Their strengths aligned with the known reasoning signatures of the source scholars.

Scholar A was consistently read as stronger in conceptual boundary control. The preserved evaluations repeatedly highlight precision in distinguishing heritage from heritage-like attachment, symbolism from public legitimacy, and participation language from actual shifts in authority. In supervisory materials, the same bot is repeatedly praised for forcing students to decide what their concept actually does, what evidence it licenses and what claims it cannot bear.

Scholar B, by contrast, was more often praised for operationalization, mediation analysis and pedagogical scaffolding. Reviewers repeatedly note the ability to translate conceptual claims into research design, to identify empirical anchors, and to track how digital infrastructures shape the visibility and social life of heritage claims.

These differences were not hard-coded as evaluation targets, they emerged from the distillation process itself. It suggests the method captured scholar-specific reasoning architectures but not just a generic ``strong academic'' style. The differential profiles were independently recovered by all three evaluation groups using different descriptive vocabularies. The convergence of direction across divergent evaluative languages constitutes strong evidence that the distillation process captured scholar-specific reasoning architectures.

\subsection{Student usability assessment (\texorpdfstring{$n = 10$}{n = 10})}

Demand-side evidence is consistent with, though not by itself sufficient to establish, a preference signal among frontier-model-using research students in the author's disciplinary network. Ten research-degree students completed the relic-bot survey after sustained use across self-selected academic scenarios. All 10 participants (100\%) answered ``yes'' to ongoing frontier-model use. These were therefore respondents comparing them to an existing baseline of regular LLM-mediated academic work. Given the small sample size, these results should be interpreted as exploratory pilot data rather than as robust inferential evidence.

Performance ratings were compressed near the top of the 7-point scale across all five Q4 dimensions. Information reliability had a mean of 6.800 with a standard deviation (s.d.) of 0.422. Innovation inspiration averaged 6.400 (s.d.\ 0.843). Academic knowledge averaged 6.600 (s.d.\ 0.516). Theoretical depth averaged 6.800 (s.d.\ 0.422). Logical rigor averaged 6.800 (s.d.\ 0.422). When averaged within participant, the composite performance score was 6.680 (s.d.\ 0.413). The distribution is heavily left-skewed with pronounced ceiling effects; interpretation of mean differences across Q4 items should therefore be cautious, and the small sample size further limits the inferential leverage of these descriptive statistics.

Confidence ratings remained near the top of the same scale. Confidence in information-reliability judgments averaged 6.700 (s.d.\ 0.483); innovation-inspiration confidence 6.500 (s.d.\ 0.527); academic-knowledge confidence 6.700 (s.d.\ 0.483); theoretical-depth confidence 6.600 (s.d.\ 0.516); and logical-rigor confidence 6.600 (s.d.\ 0.516). The composite confidence score was 6.620 (s.d.\ 0.394).

Table~\ref{tab:student-stats} reports the item-level descriptive statistics.

\begin{table}[htbp]
\centering
\footnotesize
\caption{Student survey descriptive statistics (\(n = 10\)).}
\label{tab:student-stats}
\begin{tabular}{lrrrrrr}
\toprule
Measure & \(n\) & Mean & Median & s.d. & Min & Max \\
\midrule
Q4.1 Information reliability & 10 & 6.800 & 7.000 & 0.422 & 6 & 7 \\
Q4.2 Innovation inspiration & 10 & 6.400 & 7.000 & 0.843 & 5 & 7 \\
Q4.3 Academic knowledge & 10 & 6.600 & 7.000 & 0.516 & 6 & 7 \\
Q4.4 Theoretical depth & 10 & 6.800 & 7.000 & 0.422 & 6 & 7 \\
Q4.5 Logical rigor & 10 & 6.800 & 7.000 & 0.422 & 6 & 7 \\
Q4 composite performance & 10 & 6.680 & 6.800 & 0.413 & 5.8 & 7.0 \\
Q5.1 Confidence: info reliability & 10 & 6.700 & 7.000 & 0.483 & 6 & 7 \\
Q5.2 Confidence: innovation & 10 & 6.500 & 6.500 & 0.527 & 6 & 7 \\
Q5.3 Confidence: acad.\ knowledge & 10 & 6.700 & 7.000 & 0.483 & 6 & 7 \\
Q5.4 Confidence: theor.\ depth & 10 & 6.600 & 7.000 & 0.516 & 6 & 7 \\
Q5.5 Confidence: logical rigor & 10 & 6.600 & 7.000 & 0.516 & 6 & 7 \\
Q5 composite confidence & 10 & 6.620 & 6.600 & 0.394 & 6.0 & 7.0 \\
\bottomrule
\end{tabular}
\end{table}

The usage-scenario distribution indicates that respondents turned to the scholar-bots precisely where disciplinary judgment matters most. Theory comparison and discussion was the most common scenario, selected by 9 of 10 participants (90.0\%), and it also produced one of the highest mean composite performance ratings (6.689). Learning course content was selected by 6 participants (60.0\%), academic writing polish and data-collection assistance were each selected by 5 participants (50.0\%), and literature review by 2 (20.0\%). The bots were therefore valued most in tasks that require conceptual contrast, synthesis, disciplinary framing and writing guidance.

Hypothetical stated willingness-to-pay data indicate interest beyond mere curiosity, but the small sample demands cautious interpretation and these figures should not be read as market forecasts. Stated current monthly spending on large models or related APIs had a mean of RMB 50.5 and a median of RMB 39.0, but the mean was inflated by one heavy spender (RMB 140). Hypothetical stated willingness to pay for a relic-style model had a mean of RMB 40.1 and a median of RMB 36.5. For 7 of 10 participants (70.0\%), hypothetical stated willingness to pay exceeded current monthly model spend. Hypothetical WTP is known to overstate revealed WTP; these figures should be read as directional information.

The key interpretive point is straightforward. These users were experienced users of frontier systems, yet they rated the distilled scholar-bots at high levels for precisely those academic tasks where generalized chat systems usually underperform: theory comparison, literature framing, writing structure and judgment-laden synthesis. Read cautiously and in light of the small sample, this pilot pattern is consistent with a preference signal for scholar-shaped systems among frontier-model-using research students in the author's disciplinary network.

\subsection{Capability distribution across functions}

Review and supervision remain the most stable dimensions: every preserved evaluation in those families judged benchmark attainment, and the appointment syntheses repeatedly placed both bots at Senior Lecturer level or above. Lecture was the most qualified dimension because formal Q\&A design and packaging caveats recurred across reviewers. Panel exchange was more contingent than review or supervision, but it was also more revealing. Speaking order altered visible originality and agenda-setting, while the third discussant shifted the burden from bilateral critique toward framework adaptation and infrastructural reframing. Across those changes, the bots did not merely maintain coherence; they generated real-time theoretical labor by proposing new analytic distinctions, renaming weak evaluative targets, and converting critique into alternative research architectures. Distillation therefore appears strongest where standards are stable and outputs are structured, but it is already capable of substantial real-time academic exchange.

This asymmetry is theoretically important. The distillation process does not appear to produce a substitute for intellectual invention in the strongest sense. It produces something narrower but institutionally sufficient: a system capable of applying stable standards, diagnosing argumentative weakness, reconstructing conceptual fields, sequencing revisions and guiding learning. That narrower capability is already enough to threaten a large share of the academic functions for which scholars are hired, evaluated and paid.

\subsection{Institutional-grade labour classification}

Six independent appointment-level syntheses were produced across three evaluation groups for the two primary scholar-bots. For Scholar A, recommendations ranged from Senior Lecturer through conditional Associate Professor to Professor. For Scholar B, recommendations ranged from Senior Lecturer through Senior Lecturer with strong upward indication to Associate Professor. Across all six syntheses, no evaluator placed either scholar-bot below Senior Lecturer level.

Critically, the lower bound was produced by the most evidence-strict evaluator in the pool. That evaluator's argument was not that the work lacked maturity but that appointment above Senior Lecturer would require evidence types that the protocol deliberately withheld: publication record, grant leadership, wider institutional contribution, and completed doctoral supervision. In other words, the floor was set by evidence-type constraint rather than by quality deficit.

\section{Discussion}

\subsection{Why scholarly reasoning is highly distillable}

The central empirical finding is straightforward: a scholar's public corpus can be transformed into a reusable reasoning apparatus that performs institutional academic work. The reason this is possible is that the academic publication system itself manufactures the conditions of capture.

Published papers and monographs are the most condensed and professionally ratified products. In many humanities and social science fields, especially those still structured around solo-authorship norms, the public corpus displays an individual's recurring object definitions, favored distinctions, evaluative thresholds, citation habits and refusal patterns with unusual clarity. Publication pressure rewards coherence; peer review punishes unstructured drift; careers depend on leaving a legible, citable, cumulative trail. The same institutional forces that make scholarship recognizable to peers also make it extractable to machines.

The relevant similarity threshold is therefore structural rather than theatrical. A distilled system does not need to reproduce a scholar's full inner life to be judged strikingly similar in practice. It needs to inherit the scholar's upstream architecture of judgement: how the object is defined before analysis begins, which distinctions are treated as decisive, what evidence is admissible, how objections are ranked, what theory clusters are naturally recruited, and which apparently plausible moves are refused as category errors.

\subsection{From tacit knowledge to distillable reasoning procedures}

Classical discussions of tacit knowledge emphasize the gap between what experts can do and what they can fully state \citep{Polanyi1966Tacit}. That insight remains important, but it requires disaggregation. \citet{Collins2010TacitExplicit} distinguishes three species of tacit knowledge: relational tacit knowledge, which could in principle be made explicit but has not been because no one has found reason to articulate it; somatic tacit knowledge, which is carried by bodily dispositions and cannot be fully rendered in propositional form; and collective tacit knowledge, which inheres in the social practices of a community and resists individual codification. This taxonomy reframes the distillability question. The academic publication system selectively externalizes precisely the layer of tacit knowledge that carries the greatest institutional value.

What the scholar-bots reconstruct is predominantly relational tacit knowledge, the object definitions, evaluative thresholds, diagnostic heuristics and inferential sequences that a scholar deploys routinely but rarely states as explicit rules, because peer audiences already share the disciplinary context in which those rules operate. Publication makes these procedures recoverable, because the communicative demands of peer-reviewed writing force a degree of procedural transparency that exceeds what the scholar would articulate in conversation. The somatic and collective dimensions of scholarly expertise, embodied judgment, institutional feel, the lived texture of seminar exchange, remain largely outside the distillation reach. Yet this limitation is institutionally secondary. Universities do not hire most scholars because they possess irreplaceable embodied intuition. They hire them to review, supervise, teach, assess and advise, tasks that depend heavily on the relational tacit knowledge that publication inadvertently exposes.

The result therefore reframes the complementarity debate in labor economics \citep{Autor2015Jobs,FreyOsborne2017Employment}. The usual argument holds that technology displaces routine tasks while complementing non-routine expertise. Our evidence suggests a different vulnerability class: non-routine labor can still be functionally captured when its reasoning substrate is legible, stable and textually exposed. What matters is not whether the task feels creative from the inside, but whether its institutionally rewarded execution depends on the layer of tacit knowledge that publication has already externalized.

An obvious attribution objection is that the benchmark-attaining results may reflect generic frontier-model capability rather than the distillation procedure itself. The broader project record bears directly on that concern. In addition to the scholar-bound runs reported here, the same four academic function domains were trialed with leading general chat interfaces from Gemini, ChatGPT and Claude, using their strongest generally available reasoning-oriented configurations as capability baselines. In the author's judgement, those unconstrained systems did not reach passing quality across supervision, peer review, lecture and panel tasks as a set. The present paper does not tabulate those baseline trials with the same archival fullness as the distilled systems, so the narrowest defensible claim is not that distillation explains every marginal increment of quality, but that benchmark-attaining, scholar-differentiated performance appeared only once corpus-bound reasoning constraints were introduced.

The order-reversal experiment between Rounds 1 and 2 provides additional indirect evidence. If the outputs were primarily generated by the base model's generic conversational competence, one would expect broadly similar debate patterns regardless of speaking order. Instead, evaluators reported that order systematically redistributed initiative while preserving scholar-specific reasoning signatures. That is the pattern one would expect from stable speaker-specific reasoning architectures, not from generic model fluency wearing different labels.

\subsection{The Relic condition: from encapsulation to ruin}

The Relic condition names the structural situation that emerges when public intellectual production becomes the training substrate for functionally deployable replicas. It has three features, each of which connects to a distinct strand of critical theory on technology, labor and extraction.

The first is asymmetric legibility. The scholar's reasoning system is legible in the publication record, but the downstream extraction of that reasoning is not necessarily legible to the scholar. The article is written for peers, students and institutions; it can also be read against the author by systems designed to capture its recurrent procedures. The scholar sees the act of publication. The scholar may not see the act of distillation. This asymmetry is structural, and has a precise analogue in what \citet{Pasquale2015BlackBox} calls the black box society: institutions and platforms that benefit from transparency in one direction (the scholar's public output) while maintaining opacity in the other (the extraction pipeline, the downstream deployment, the commercial logic). In the Relic condition, the asymmetry is doubled: the scholar's reasoning is maximally transparent by professional obligation, while the extraction apparatus that feeds on that transparency is maximally opaque by commercial design.

The second is involuntary contribution. Academic labor is already organized under publish-or-perish conditions. Scholars cannot simply withhold their reasoning architectures from public circulation without professional cost. Every article that sharpens a concept, stabilizes a method or refines an evaluative distinction may simultaneously improve the next generation of extractive artifacts. The scholar is compelled to feed the system that may reduce the need for the scholar's own future labor. Couldry and Mejias \citeyearpar{CouldryMejias2019Costs} argue that the systematic appropriation of human life-activity through data relations constitutes a new form of colonialism, one that seizes not territory but the capacity to render life into data for value extraction. The Relic condition extends this diagnosis to a domain Couldry and Mejias do not specifically address: the colonization of scholarly reasoning. Where data colonialism captures behavioral surplus from everyday digital life, reasoning colonialism captures cognitive surplus from professional intellectual production. The structural logic is the same: an existing social relation (in this case, publication for peer communication) is annexed as raw material for an extractive apparatus that its originators neither authorized nor control. The difference is that scholarly colonization is doubly involuntary. The scholar is compelled to publish by career structure, and the publication is compelled to be legible by peer review norms. Both compulsions are prerequisites of academic survival, and both simultaneously prepare the corpus for extraction.

The third is sufficiency without equivalence. A Relic does not need to become the scholar. It needs only to be useful enough for institutions to substitute it into key functions. Reviewing, supervisory diagnosis, lecture preparation, curriculum support, consultation and first-pass conceptual triage all operate under thresholds of adequacy, not thresholds of metaphysical equivalence. Once a distilled artifact is good enough to satisfy those thresholds cheaply and at scale, the question of whether it fully equals the scholar becomes economically secondary. This mechanism is not unique to AI-driven displacement; it recapitulates a logic that Marx \citeyearpar{Marx1844Economic} identified in industrial production, where the replacement of skilled artisanal labor did not require machines that matched the artisan's full capacity, only machines that produced output adequate to the market's operative standard. The sufficiency threshold in academic labor is set not by philosophical judgment about the nature of thought, but by institutional audit cultures, workload models and cost-per-output calculations that are already biased toward standardized, scalable delivery.

These three conditions generate a self-reinforcing dynamic. First comes the distillation race: once the method is known and the demand signal is visible, every field's most legible scholars become candidates for corpus distillation. Then comes the SOTA game: once multiple scholar-Relics coexist, they become rankable. Which one gives the sharpest supervision, the most actionable review, the clearest lecture? The SOTA game does not necessarily reward the most profound or original thinker. It rewards the thinker whose public reasoning system is most extractable, most stable and most deployable. Modular, highly taxonomized, cross-text-consistent scholars will tend to yield stronger Relics than thinkers whose work is deliberately tensioned, self-revising or resistant to codification. In that sense, the SOTA game selects for distillability over originality. Finally comes relicification: every new publication contributes to the refinement of an artifact that can perform pieces of the scholar's labor more cheaply than the scholar can. Returns to innovation are split off from the innovator and transferred to the deployers of the Relic: platforms, labs, publishers, universities, firms. This creates what we call innovator nihilism. If every refinement of my reasoning system increases the fidelity of my own replacement, then publication starts to undermine the incentive structure that made publication possible. At enough scale, a system built on extraction begins to consume the motivational substrate of innovation itself.

Marx's \citeyearpar{Marx1844Economic} discussion of alienation turns on the worker's loss of relation to the product, the process and the conditions of species-being. The Relic condition is a specifically intellectual form of that process: the scholar's public product becomes alienable not only in the ordinary publishing sense but in a second-order functional sense, as a reusable apparatus separable from the scholar and mobilizable against the scholar's future labor market position. What distinguishes this from classical alienation is not the structure of the process but the specificity of the object. Industrial alienation separates the worker from a material product. Scholarly alienation separates the thinker from a reasoning architecture, an object that is simultaneously the scholar's professional identity, career asset and, once extracted, competitive replacement.

Stiegler's \citeyearpar{Stiegler2009TechnicsTime2} account of technics as tertiary retention clarifies the storage form at stake. For Stiegler, tertiary retentions are externalized memory supports, from writing to recording to computation, that carry past experience forward in forms independent of the remembering subject. Distilled reasoning architectures are a new species of tertiary retention: they preserve not declarative content but procedural logic, the sequence of operations by which a scholar constructs objects, licenses evidence, ranks objections and refuses false solutions. What is retained is not what the scholar said but how the scholar thinks, in a portable format that can be replayed without the scholar's participation and against the scholar's interests. Where Stiegler's earlier examples of tertiary retention, such as writing, television and digital recording, primarily concerned the externalization of narrative and perceptual memory, the Relic represents the externalization of inferential procedure itself. The scholar's reasoning is no longer merely recorded; it is operationalized as an executable constraint on a general-purpose computational system.

The Relic condition also differs from Pasquinelli and Joler's \citeyearpar{PasquinelliJoler2021Nooscope} nooscope framework in specificity and target. The nooscope describes AI as a general instrument of knowledge extractivism operating on collective data labour. The Relic condition narrows this to a named mechanism: the distillation of individual reasoning architectures from identifiable scholarly corpora. Where the nooscope foregrounds the aggregate statistical logic of training on massive datasets, the Relic condition foregrounds the targeted logic of corpus-level reconstruction, in which a single scholar's publicly sedimented reasoning becomes the extraction target. This specificity matters because it identifies the publication system itself, rather than data accumulation in general, as the extraction infrastructure, and because it assigns vulnerability not to anonymous data subjects but to identifiable knowledge workers whose professional output is the object of capture.

\subsection{The AI--capital nexus and accumulation by dispossession}

While the Relic condition describes the structural possibility of scholarly reasoning capture, its translation into actual institutional displacement requires an additional force: the capital interests that make such displacement rational at scale. The AI--capital nexus arises when two forces meet: AI systems that can cheaply execute extracted reasoning, and institutional actors that benefit from substituting those systems for slower, more expensive human labor.

Higher education is a particularly acute site because it sits at the intersection of academic capitalism, public-knowledge norms and labor-market austerity \citep{SlaughterRhoades2004AcademicCapitalism,Gill2009BreakingSilence}. Universities are already under pressure to scale teaching, reduce staffing costs, expand monitoring and convert scholarly work into auditable output. A scholar-bot able to support supervision, writing feedback, theory explanation or peer review is therefore not merely a research curiosity. It is an immediately legible managerial instrument. The same is true of publishers and platforms. If reviewer labor, editorial triage or pedagogical support can be partially replaced by scholar-shaped systems, cost savings will pull deployment forward long before philosophical questions about equivalence are settled.

\citet{Srnicek2017PlatformCapitalism} argues that the dominant tendency in digital capitalism is toward the platform as the organizational form that extracts value by intermediating between different user groups while retaining control over the infrastructure of interaction. Scholar-Relics, once productized, follow this logic precisely. A platform offering distilled scholarly reasoning as a service positions itself between students seeking supervision, institutions seeking scalable review capacity, and the scholars whose extracted architectures power the system. The platform captures the intermediary rent; the scholar receives nothing. Crucially, the platform logic also explains why scholar-Relics will tend to become proprietary assets rather than public goods. The competitive advantage of a platform lies in exclusive access to the reasoning architectures it has captured, not in their open circulation. This is why voluntary frameworks that appeal to goodwill are unlikely to be sufficient: the economic logic of platform capitalism actively rewards enclosure.

Harvey's \citeyearpar{Harvey2003NewImperialism} concept of accumulation by dispossession provides the political-economic frame for this process. Harvey argues that capital accumulation increasingly proceeds not through expanded reproduction alone, but through the seizure and privatization of previously common or public assets: land enclosures, intellectual property extensions, the commodification of cultural forms, the privatization of public services. Scholarly reasoning distillation is a new instance of this logic. The public scholarly record, built through public funding, open-access mandates and publish-or-perish compulsion, constitutes a knowledge commons. Corpus-targeted distillation converts portions of that commons into proprietary deployable assets without compensation, consent or even notification. The dispossession is structurally invisible because it does not remove the original publications from circulation. It extracts their operative value while leaving the textual shell intact. The scholar's corpus remains publicly available; its functional content has been enclosed.

In this respect the loop extends the extractive logic described in surveillance-capitalism accounts \citep{Zuboff2019Age}, where behavioral surplus becomes a target of systematic capture and redeployment for institutional advantage. But it differs in a critical respect. Zuboff's behavioral surplus is generated as a by-product of digital activity that users would engage in anyway. Scholarly reasoning surplus is generated as the primary product of professional intellectual labor: the article, the monograph, the review. The scholar is not merely surveilled while doing something else. The scholar's most deliberate, most professionally rewarded output is itself the raw material of extraction. This makes the Relic condition not a side effect of surveillance capitalism but a direct extension of its logic into the core of knowledge production.

These forces converge to produce what we term the AI--capital predation loop: AI captures cognitive labor; capital deploys captured labor at lower cost; the original laborers lose institutional justification; incentives to produce new high-quality cognitive labor weaken; and the innovation base on which AI feeds begins to decay. The loop is predatory because it transfers value from originators to deployers while eroding the conditions under which new originators can emerge. It is also self-undermining: a predation loop that degrades the quality of the knowledge base it feeds on will eventually degrade the quality of its own outputs. But the temporal mismatch between short-run cost savings and long-run knowledge degradation means that market incentives will drive deployment well past the point at which the damage becomes irreversible.

\subsection{Day 0}

The correct temporal frame for this problem is Day 0. Day 0 is the earlier window in which technical capability is already sufficient to make the threat credible, but deployment has not yet fully hardened into infrastructure. The evidence reported here places scholarly reasoning distillation squarely in that window. The method uses public corpora, no domain fine-tuning and inference-time constraints rather than bespoke model retraining. The engineering distance from demonstration to productization is therefore small.

Waiting for mass deployment before responding would repeat a familiar regulatory failure pattern. By the time institutions ask whether reasoning distillation requires consent, compensation, auditing or rights protection, the artifacts will already be in circulation, benchmarked and normalized. Protective action is cheaper before ranking markets, product ecosystems and institutional workflows lock in around the Relic. The Day 0 argument does not require certainty about the scale or speed of future deployment. It requires only that the demonstrated capability, combined with visible institutional incentives for cost reduction, makes the risk of unregulated extraction credible enough to warrant precautionary intervention.

The reflexive risk of warning itself must also be acknowledged. Publishing this paper participates in the dynamic it describes. A public articulation of scholarly distillability contributes, at the margin, to the legibility that makes distillation cheap. The strategy adopted here separates conceptual from operational disclosure: the reasoning-system framework, extraction logic at the level of layer definitions, and evaluation outcomes are reported in full, while the specific prompts, skill files and reproducible pipeline artefacts are withheld so that publication does not itself function as a distillation tutorial. This is an imperfect compromise. It remains defensible only while the gap between a warning existing and the warning being implementable at scale is still wide enough to be policy-actionable.

\subsection{Toward a protective framework}

The immediate policy question is whether the reasoning architecture embedded in public scholarship should remain wholly unprotected once extraction becomes technically trivial.

At the publication-system level, one question is whether intellectual property doctrine can or should recognize reasoning fingerprints as distinct from surface text. Current regimes are better at protecting textual expression than structural intellectual patterns, but the results here suggest that the economically decisive layer may be the latter rather than the former. Publishers, repositories and indexing infrastructures may need explicit policies governing automated corpus distillation targeted at identifiable individuals. Floridi and colleagues \citeyearpar{FloridiEtAl2018AI4People} have argued that ethical AI governance must rest on principles of autonomy preservation and informed consent; the involuntary-contribution mechanism identified here represents a direct violation of both principles, since scholars are neither informed of nor able to consent to the extraction of their reasoning architectures under current publication norms.

At the AI-governance level, developers should be required to disclose whether individual reasoning-system distillation from public corpora was used in model products or product features. Distillation audits, provenance disclosures and compensation frameworks are obvious starting points. So too are restrictions on deploying scholar-targeted systems in evaluative institutional contexts without consent from the individuals whose reasoning architectures were extracted. UNESCO's \citeyearpar{UNESCO2023Guidance} recent guidance on generative AI in education and research emphasizes the need for human-centered governance, transparency and policy intervention before educational systems are reorganized around convenience alone. The issue raised here is more specific but entirely consistent with that broader governance agenda: if publication corpora become extraction targets, then the governance of AI in research must extend beyond questions of AI-generated content to questions of AI-captured reasoning.

At the academic-community level, the challenge is strategic. The point is to identify which dimensions of scholarly life must remain human because once they are reorganized around extractive replay, the incentive to produce new thinking degrades. Doctoral education, mentorship, peer evaluation and scholarly publication all require rethinking under conditions where publicly stabilized reasoning can be cloned into low-marginal-cost artifacts. The Relic condition suggests that the greatest long-run risk is not the automation of any single academic function, but the slow degradation of the knowledge-production ecology that sustains all of them.

\section{Conclusion}

This paper has reported a small-scale but high-density demonstration that the reasoning systems of established humanities and social science scholars can be distilled from their published corpora alone and deployed as functional academic labour. The resulting scholar-bots were judged benchmark-attaining across review, supervision, lecturing and panel exchange by independent senior academics, and classified at Senior Lecturer level or above in all six appointment-level syntheses. Three independent evaluation groups consistently distinguished two different scholarly personalities in the outputs.

These findings do not establish that all scholars are equally vulnerable, that distillation at scale is imminent, or that institutional substitution is inevitable. They establish something narrower but consequential: that the technical threshold for scholarly reasoning capture has already been crossed under ordinary publication conditions, using public corpora, no domain fine-tuning and modest engineering effort. The Relic condition names the structural predicament that follows from this threshold crossing.

The window for protective action is the present. Disclosure requirements, consent frameworks, compensation mechanisms and deployment restrictions are all technically feasible and politically achievable while scholarly reasoning distillation remains a demonstrated capability rather than an entrenched infrastructure. That window is narrowing. The evidence reported here is intended to ensure it does not close unnoticed.

\section{Methods}

\subsection{Study design}

The study tested whether scholarly reasoning distillation from published corpora could produce functionally deployable scholar-bots capable of performing core academic tasks at benchmark-attaining quality. The design combines corpus-based reconstruction of reasoning systems, task-based generation through a local Codex agent using a GPT-5.4 base model, expert evaluation of generated outputs and student usability assessment after sustained applied use.

\subsection{Scholar selection and corpus construction}

The two target scholars were selected because they were leading figures in related areas of critical heritage studies yet displayed different theoretical emphases, different writing textures and different forms of disciplinary intervention. Both had sufficiently large public corpora to support full local processing. To reduce exposure risk in the manuscript, we refer to them as Scholar A and Scholar B throughout the main text and supplementary materials.

Scholar A's corpus was processed as 68 analytical units covering approximately 1,742 analytical pages: 8 books, 8 chapters and 52 papers. Scholar B's corpus coverage report records 35 fully processed local corpus items: 21 papers, 8 long-form works and 6 chapters. The corpus-only rule was strict. No web search, biographical supplementation or external knowledge retrieval was used to define the distilled reasoning systems.

\subsection{Eight-layer reasoning extraction}

Both scholar-bots were distilled with the same extraction logic. The method targeted eight layers of scholarly reasoning: ontological features, conceptual features, analytical operations, evaluative features, intertextual features, rhetorical features, boundary features and diachronic features.

Candidate features were retained only when they appeared across at least two independent texts or at least three independent argumentative contexts. Monographs, methodologically central texts and richly evidenced papers carried greater interpretive weight than translations, near-duplicates, editorials or short notices.

\subsection{Nine-module skill architecture}

The extracted reasoning layers were converted into a nine-module executable skill architecture: scope, activation, ontological, procedural, evaluative, intertextual, rhetorical, refusal and provenance/evolution. The architecture was designed to target the level at which scholar-specific similarity actually arises. Similarity in this system does not come primarily from lexical mimicry, but comes from reproducing the order in which a scholar constructs the object, licenses evidence, ranks objections, recruits theory and decides what should be refused.

Table~\ref{tab:modules} summarizes the module functions.

\begin{table}[htbp]
\centering
\caption{Nine-module skill architecture.}
\label{tab:modules}
\begin{tabular}{p{1.4in}p{4.4in}}
\toprule
Module & Function \\
\midrule
Scope & Defines corpus boundaries, time limits and non-targets \\
Activation & Governs when the skill engages, how uncertainty is marked and when it exits \\
Ontological & Reconstructs what the object of analysis is before problem-solving begins \\
Procedural & Specifies the scholar's recurrent analytical operations \\
Evaluative & Encodes thresholds for strong and weak argument \\
Intertextual & Reconstructs citation clusters, citation functions and inferential bridges \\
Rhetorical & Preserves reasoning rhythm and language organization \\
Refusal & Encodes anti-patterns, boundaries and limiting conditions \\
Provenance/Evolution & Records formation logic, confidence and internal development across the corpus \\
\bottomrule
\end{tabular}
\end{table}

\subsection{Base model}

The scholar skills were deployed on a GPT-5.4 model in a very-high-reasoning configuration. No domain-specific fine-tuning was performed. The skills operated as inference-time constraints that reshaped the model's reasoning pathway without altering model weights.

\subsection{Experimental tasks}

The scholar-bots were evaluated on four academic function dimensions: doctoral supervision, peer review, lecturing, and panel discussion. For supervision, both bots independently annotated a real early-stage doctoral proposal. For peer review, both bots reviewed the same self-authored, field-relevant manuscript. For lecturing, each bot generated course-specific lecture scripts. For panel discussion, the archive preserves three linked rounds built on the same confirmation-slide deck, including order-reversal and a three-way stress test.

\subsection{Expert evaluation protocol}

Three senior academics independently evaluated the generated outputs. The protocol was rubric-guided but not perfectly standardized in report format. Peer-review evaluations focused on standard sense, proportionality, defensibility, actionability and consistency. Supervision evaluations focused on diagnostic accuracy, priority ordering, feasibility judgment, developmental awareness and independence orientation. Lecture evaluations focused on accuracy, structure, hierarchy, learnability and Q\&A robustness. Panel evaluations focused on rigor, evidence, responsiveness, clarity and originality, using 10-point scales.

Separately from the three-evaluator protocol, the broader archive includes one source-scholar inspection of the corresponding distillate. This occurred in a formal face-to-face academic discussion.

\subsection{Student usability survey}

Ten research-degree students completed the relic-bot usability survey after using the system across one or more academic scenarios. The student-use phase was conducted on a workstation running a Codex agent loaded with three scholar skills and exposed through a lightweight chat window. Students with similar disciplinary background were invited either in person or via video call to test the system freely, but they were explicitly instructed to push the interaction toward genuine academic questions in the field.

All testers were personally known to the author. Rather than masking this, the study records it as part of the protocol. The working assumption was that acquaintance could increase the reliability of confidence judgments by making task expectations, domain standards and the intended seriousness of the exercise maximally explicit. That assumption should not be treated as a neutrality guarantee: the same social proximity also reduces anonymity and limits generalizability beyond the author's immediate disciplinary network.

The survey asked whether participants already used frontier models, their current monthly spending, the academic scenarios in which they used the relic-bots, five 7-point performance ratings (information reliability, innovation inspiration, academic knowledge, theoretical depth and logical rigor), five 7-point confidence ratings for those judgments and willingness to pay for relic-model subscription access. The survey was analyzed descriptively.

\subsection{Ethics, reflexivity, and dual-use positioning}

This study involves a deliberate ethical tension that the design makes explicit. No consent was sought from the distilled scholars because the research question is precisely whether publicly circulated scholarly corpora already expose reasoning architectures to third-party extraction without further authorization. This approach parallels responsible-disclosure traditions in computer security research.

To discharge the obligations created by that stance, the study adopts four commitments. First, the identities of the distilled scholars are withheld. Second, the full extraction prompts, skill-module files and reproducible pipeline artefacts are not released publicly. Third, if any scholar whose corpus was used requests destruction of the local skill artefacts, the author will comply. Fourth, no distilled scholar-bot has been or will be publicly deployed, productized or made accessible outside the closed evaluation described here.

The doctoral proposal, peer-review manuscript and panel-deck materials used as evaluation inputs were all authored by the present author, who self-authorizes their use in this study. Student participants were informed that they were taking part in an academic usability assessment and that only aggregated, anonymized responses would be reported. No formal ethics approval number is currently recorded in the project archive. The student survey was conducted as a low-risk usability assessment under the author's institutional guidelines; if a retrospective approval or exemption identifier becomes available, it will be added to subsequent versions. All participants provided informed consent to participate.

\subsection{Limitations}

The study has seven main limitations. First, the two primary scholar corpora were not recorded with identical unit conventions, so corpus scale is similar in ambition but not perfectly symmetrical in accounting. Second, the panel archive remains heterogeneous in format across rounds. Third, expert reports vary in form. Fourth, the student-usability cohort was socially proximate to the author, interacted through a guided local Codex interface, and comprised only 10 participants, which limits statistical power and generalizability. Fifth, the distillation reconstructs public intellectual practice, not the private totality of a scholar's mind. Sixth, the paper deliberately withholds full extraction prompts, skill files and reproducible pipeline artefacts for dual-use reasons, which constrains exact reproducibility. Seventh, the student survey sample is small (\(n = 10\)) and drawn from a single disciplinary network; the survey results should therefore be treated as exploratory pilot data and cannot be robust inferential evidence.

\section*{Supplementary Information}

The Supplementary Information file accompanies this paper as a single document organized into numbered sections (S1--S32). It provides the technical background, de-identified task dossiers, and representative output evidence needed for peer review. The file is structured to make the technical object of the paper legible without disclosing sensitive raw documents, copyrighted corpora, or directly replicable extraction artefacts.

Sections S1--S5 document the runtime system, corpus coverage, extraction protocol, skill architecture, and validation logic. Sections S6--S7 describe the experimental task matrix and student-use environment. Sections S8--S17 provide de-identified input and output dossiers for each task family (supervision, peer review, lecturing, and panel discussion), with identifying details of the author's unpublished manuscripts redacted to protect pre-publication interests. Section S18 provides a representative de-identified output evidence bank organized by task and scholar-bot. Section S19 specifies permissions, redaction boundaries, and release constraints. Section S20 maps the relationship between the main paper, Supplementary Information, and planned Extended Data. Sections S21--S26 contain extended de-identified expert-evaluation records across all three evaluator groups for both primary scholar-bots, limited to general methodological observations and evaluator judgments that do not disclose the specific topic, argument structure, or findings of the unpublished manuscript under review. Section S27 summarizes reliability-by-density across the evaluation archive. Sections S28--S31 provide panel score-file exemplars for all three rounds. Full panel transcripts are held in the project archive and can be provided to reviewers on request in de-identified form.

The following content is excluded from the Supplementary Information: (a) specific review comments from the scholar-bots on the author's unpublished manuscript that would reveal its subject matter, conceptual framework, or empirical findings; (b) verbatim or near-verbatim reproduction of the bots' peer-review reports on the manuscript; (c) any material from which the manuscript's core claims could be reconstructed prior to its independent publication; and (d) the full extraction prompts, skill-module files, and reproducible pipeline artefacts, which are withheld for dual-use reasons as described in Methods 5.9.

\bibliographystyle{plainnat}
\bibliography{references}

\end{document}